# Field Scanner Design for MUSTANG of the Green Bank Telescope

Jingquan CHENG[1*], Yang LI[2], Xinnan LI[2], Brian MASON[1]

[1] *National Radio Astronomy Observatory, Virginia 22903, USA;*
[2] *Nanjing Institute of Astronomical Optics and Technology, Jiangsu 210042, China*

MUSTANG is a bolometer camera for the Green Bank Telescope (GBT) working at a frequency of 90 GHz. The detector has a field of view of 40 arcseconds. To cancel out random emission change from atmosphere and other sources, requires a fast scanning reflecting system with a few arcminute ranges. In this paper, the aberrations of an off-axis system are reviewed. The condition for an optimized system is provided. In an optimized system, as additional image transfer mirrors are introduced, new aberrations of the off-axis system may be reintroduced, resulting in a limited field of view. In this paper, different scanning mirror arrangements for the GBT system are analyzed through the ray tracing analysis. These include using the subreflector as the scanning mirror, chopping a flat mirror and transferring image with an ellipse mirror, and chopping a flat mirror and transferring image with a pair of face-to-face paraboloid mirrors. The system analysis shows that chopping a flat mirror and using a well aligned pair of paraboloids can generate the required field of view for the MUSTUNG detector system, while other systems all suffer from larger off-axis aberrations added by the system modification. The spot diagrams of the well aligned pair of paraboloids produced is only about one Airy disk size within a scanning angle of about 3 arcmin.

**antenna, optics, off-axis, linear astigmatism, coma, reflector scanner**

# 1 Introduction

MUSTANG (Multiplex SQUID TES Array at Ninety GHz) is a 90 GHz bolometer with a large focal plane array of 64 detector units designed for the Green Bank Telescope (GBT) [1]. Equipped with extremely sensitive superconductor transition edge sensors (TES) and superconductor quantum interference devices (SQUIDs), MUSTANG forms diffraction limited images at 90 GHz frequency and is suitable for a wide range of astronomical observations. It has a 40 arcsecond field of view. For cancelling out noises from the atmosphere and electronics, Mason proposed a fast scanning/chopping device over a 3 arcminute range for the detector system. With this chopping/scanning system, noises common to many pixels are eliminated while astronomical signals are extracted with no loss in efficiency.

The GBT off-axis optics is optimized without the linear astigmatism aberrations which are dominant in an off-axis system. The system details were discussed in the GBT memo 155 by Norrod and Srikanth [2]. Moving the subreflector away from its ideal position or adding image transferring mirrors in the system, usually destroys this optimized condition, resulting in ugly beams and images with strong linear astigmatism as well as coma. At the same time, the pulse tube operation of the MUSTANG cooling system degrades when tipped from the vertical angle, producing operational difficulties when the detector is located in the GBT Gregorian turret. Therefore moving the MUSTANG device out from the GBT turret and mounting it in a fixed location above the platform would yield significant operation improvement. For these two reasons, a well-behaved chopping mirror and an optimized imaging transferring system is urgently required to allow MUSTUNG device's fast scanning over a larger field angle range.

A general review of the aberrations and polarization of an off-axis optical system was made by Cheng et al [3] and Cheng [4]. More detailed discussions were given by Mizugutch [5], Dragone [6], Rusch et al [7], Noethe et al [8], Chang and Prata [9], and Chang [10]. Certain conditions have to be met for an optimized off-axis system to avoid unacceptable linear astigmatism and cross polarization in an off-axis system,. Otherwise, an off-axis optical system may have zero field of view which allows the only observation at one focal point. The required condition as shown later is only for a two-mirror system, not a single mirror. In this paper, ray tracing examples of subreflector chopping and of flat mirror choppers with different image transferring systems are presented. In the end, a flat mirror chopper with a well aligned coma-free and linear astigmatism free Czeny-Turner arrangement of two face-to-face paraboloid reflectors is presented. At the same time, the focal plane of the paraboloid mirror is nearly parallel to the GBT focal plane for avoiding off-focus aberrations. Therefore, the image quality of the designed chopper meets the fast scanning and field of view requirements.

# 2 Field of view for an off-axis optical system

For axial symmetric optical systems, the major aberrations include spherical aberration, linear coma, quadratic astigmatism, and field curvature if scale change is ignored. Any axial symmetrical surface will have a fixed coma, astigmatism, and field curvature, but a conic constant related spherical aberration. A concave paraboloidal surface is free from spherical aberrations. The above statement fits a surface where its entrance pupil is located at the surface position. If a reflecting mirror system has more than one surface, one surface may be away from its entrance pupil. As the entrance pupil moves, additional aberrations will be produced. In this case, spherical aberration will produce coma, and spherical aberration and coma will produce stigmatism. There are no constant coma and linear astigmatism in the system. However, as reflecting components of an axial symmetrical system shift away from their original axial symmetrical positions, pointing change, constant coma, as well as linear astigmatism are produced. The constant coma fills the whole field of view and the linear astigmatism increases with the field angle. The constant coma and linear astigmatism can only be canceled by shifting or rotating another component away from the symmetrical axis.

An off-axis system has an obvious advantage over an axial symmetrical one as it provides a clear and blockage free aperture field. For radio and infrared telescopes, it avoids much of the background noise. However, an off-axis optical system has complex aberrations and a small or zero field of view. Chang [10] produced expressions for the optical path length function for an off-axis optical system. From these expressions, an off-axis system contains all kinds of aberrations, with linear astigmatism and third order coma being the dominant ones. The strong astigmatism of the off-axis system is also closely related to the cross-polarization of radio telescopes. With these complex aberrations, an off-axis system may have a small or zero field of view.

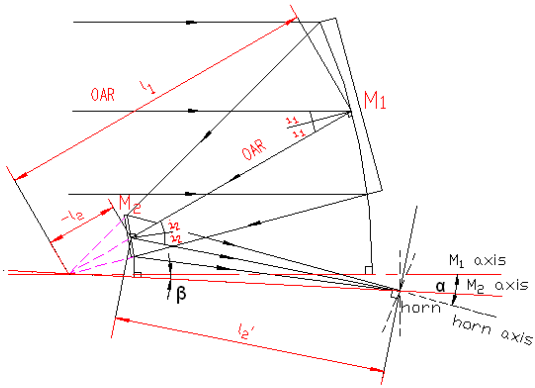

**Figure 1** System parameters of a linear astigmatism free off-axis two-mirror system. Note: OAR is the optical axis ray.

In the design of radio telescopes, efforts are made to suppress the cross-polarization within an off-axis system. Rusch et al [11] provided rigorous formula for an off-axis system based on the equivalent axial symmetrical paraboloid system. The condition to achieve a cross-polarization free system is:

$$\tan\alpha = \frac{|1-e^2|\sin\beta}{(1+e^2)\cos\beta - 2e} \text{ or}$$

$$\tan\frac{\alpha}{2} = \frac{e+1}{e-1}\tan\frac{\beta}{2} \quad (1)$$

where $\alpha$ is the angle between the horn axis and the subreflector (M2) axis, $\beta$ is the angle between the subreflector axis and the primary mirror (M1) axis, and $e$ is the subreflector (M2) eccentricity (Figure 1). In fact, the cross-polarization of linear excitation and beam displacements in circular polarized excitation are directly related to the astigmatism in geometrical optics. In 2005, Chang and Prata discussed the field of view of an off-axis system based on the pure geometrical aberration theory. They derived an important formula for cancelling out linear astigmatisms of the system. The formula derived is, in fact, identical to the earlier Rusch's formula shown in Equation (1) as:

$$\frac{l_1}{R_1}\sin 2i_1 = \frac{l_2}{R_2}\sin 2i_2 \quad (2)$$

where $l_1$ and $l_2$ are distances (Figure 1) from the common focus to the primary and the subreflector at the incident points of the optical axis ray (OAR), $i_1$ and $i_2$ the incident angles of the OAR at the primary and subreflector and $R_1$ and $R_2$ the vertex radius of primary and subreflector mirrors.

The above expression also has other forms. An alternative form of this formula was derived by Dragone [6] as:

$$m_2 \tan i_1 = (m_2 \pm 1)\tan i_2 \quad (2)$$

where $m_2$ is the magnification factor of the subreflector. The plus sign in the formula is for a Gregorian system, while the minus one is for a Cassegrain system.

Off-axis systems, once satisfying the above condition, will be free from linear astigmatism and cross polarization. The dominant remaining aberration is the linear coma in an optimized system. The GBT optical system is an optimized one [12,13]. At its Gregorian focus, it is free from linear astigmatism, producing a useful field view up to 4 arcmin. The ray tracing spot diagrams over a 4 arcmin field of view is shown in Figure 2. In the figure, from left to right and from top to bottom, the image spots are at (0, 0), (0, 2 arcmin), (0, 4 arcmin), (0, -2 arcmin), (0, -4 arcmin), (-2 arcmin, 0), (-4 arcmin, 0), (2 arcmin, 0), and (4 arcmin, 0) positions respectively. The circle diameter is 7.72 mm, the size of the Airy disk at 90 GHz.

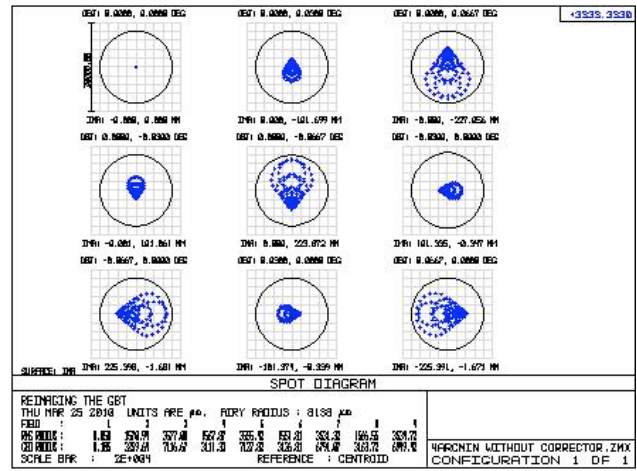

**Figure 2** Ray tracing results of the GBT optics within a 4 arcmin field of view (Circles are Airy disk at 90 GHz).

## 3 Scanner design for the GBT optics

For axial symmetric telescopes, scanning a small area of sky can easily be realized by either moving the whole telescope or chopping the subreflector mirror. Unfortunately, the GBT is a heavy and big structure, and moving the whole telescope is extremely slow. Efforts had been taken in chopping the GBT subreflector mirror. As it is discussed, any movement of the subreflector mirror away from its idea position results in ugly beams full of astigmatism and coma. For the GBT, as the subreflector tilts 0.12 degrees away from its ideal position, the beam will move 1 arcmin on the sky. The reduction of Strehl ratio is about 50% as shown in Figure 3[1]. The image spread on the MUSTANG receiver increases from about 8"×8" in the focal center (telescope focal point) to about 15"×15" at 1 arcmin away from the field center. A fat beam shape in one direction is produced by using the subreflector scanning from the focal center to 1 arcmin away from the center. The beam has a strong asymmetric, ellipse shape as the linear astigmatism dominants the distorted off-axis optical system.

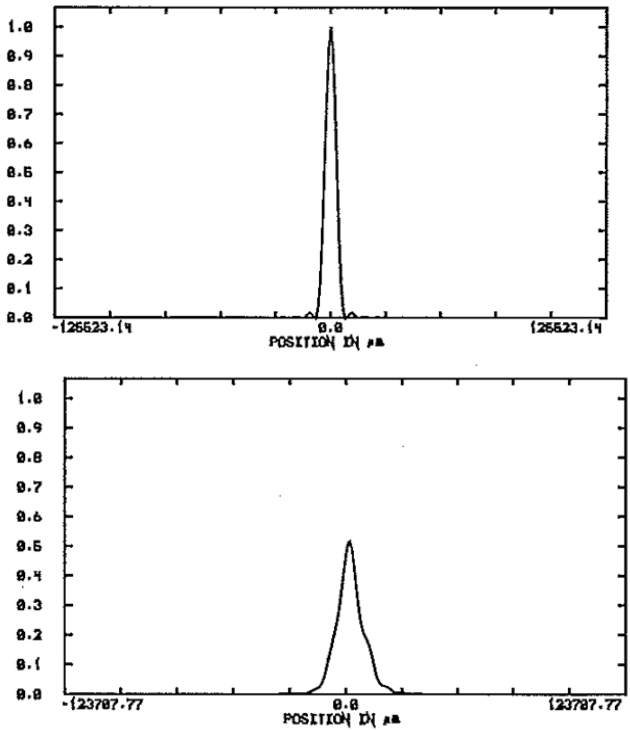

**Figure 3** *Upper*: Beam pattern of the GBT optics at the focal center. *Bottom*: Beam pattern in one direction using subreflector scanning at 1 arcmin away from the focal center (Dicker).

The alternative way to quickly scan the field is to insert a rotating mirror near the focus with its axis slightly away from the mirror surface normal. When the mirror rotates around this axis, the beam will move back and forth around two end positions in the sky. For this system, an elliptical mirror can be used to redirect the beam from the focus to a new position. In theory, a perfect image in the focus can be transferred to the new focus without any distortion. However, the added surface itself is another off-axis system which does not satisfy the linear astigmatism free condition. Therefore, any off-focus image will result in excessive linear astigmatism and constant coma.

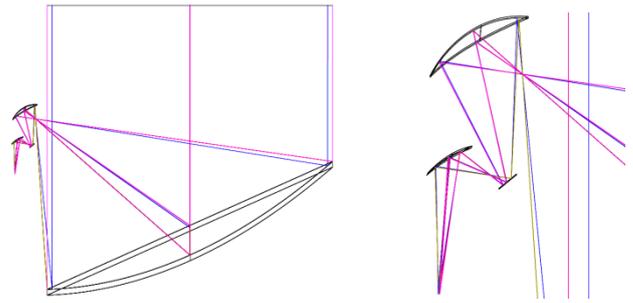

**Figure 4** Optical arrangement using a flat chopper and an ellipse reflector mirror.

Figure 4 shows both the proposed chopping flat mirror and the ellipse mirror near the GBT focal position. Figure 5 is the ray tracing spot diagrams at the focal center and about 3 arcmin away from it, both for the "flat chopper" case. From the upper spot diagrams, the image is perfect over 23 arcsec field of view at the focal center. From left to right and top to bottom, the spot diagrams are for (0, 0.0064 deg), (0.0064 deg, 0), (0,0), (-0.0064 deg, 0), and (0,-0.0064 deg) field angles. However, when the sky is 3 arcmin away from the field center, the image quality becomes very poor with strong inherited linear astigmatism covering a few Airy disk areas. In the bottom spot diagrams, from left to right and top to bottom are for (0, -0.0436 deg), (0.0064 deg, -0.05 deg), (0, -0.05 deg), (-0.0064 deg, -0.05 deg), and (0, -0.0564 deg) field angles.

By adding two curved mirrors to the image transferring system, the linear astigmatism free conditions can be satisfied. Therefore, both the GBT and the added optical system are optimized off-axis systems. Following this approach, a Czerny-Turner arrangement of two face-to-face paraboloid reflectors can be used [14]. These two paraboloids have the same f-ratio and they form a coma-free, linear astigmatism free, optimized off-axis system. Figure 6 is such an arrangement. From the figure, it can be found that the focal plane of the added image transferring system has a larger tilting angle relative to the focal plane of the original GBT system, so that the off-focus image of the GBT optics may be away from the focal plane of the added system. This will introduce axial aberrations to the input image in the image transfer system. Figure 7 shows the spot diagrams of the system around the focal center and 3 arcmin away from the focal center. From the upper spot diagrams, the image is perfect only in the focal center. The rest of the images in the field are all larger than the Airy disk. From left to right and top to bottom the spot diagrams are for (0, -0.0064 deg), (-0.0064 deg, 0), (0, 0), (0.0064 deg, 0), and (0, 0.0064 deg) field angles. When the beam is 3 arcmin away from the focal center, the image quality in the bottom spot diagrams is even poorer with strong axial aberrations related linear astigmatisms which cover many Airy disk areas. The spot diagrams from left to right and top to bottom are for (0,

-0.0436 deg), (0.0064 deg, -0.05 deg), (0, -0.05 deg), (-0.0064 deg, -0.05 deg), and (0, -0.0564 deg) field angles. The transferring of the GBT image field is poor because the focal plane is tilted away from the focal plane of the GBT telescope.

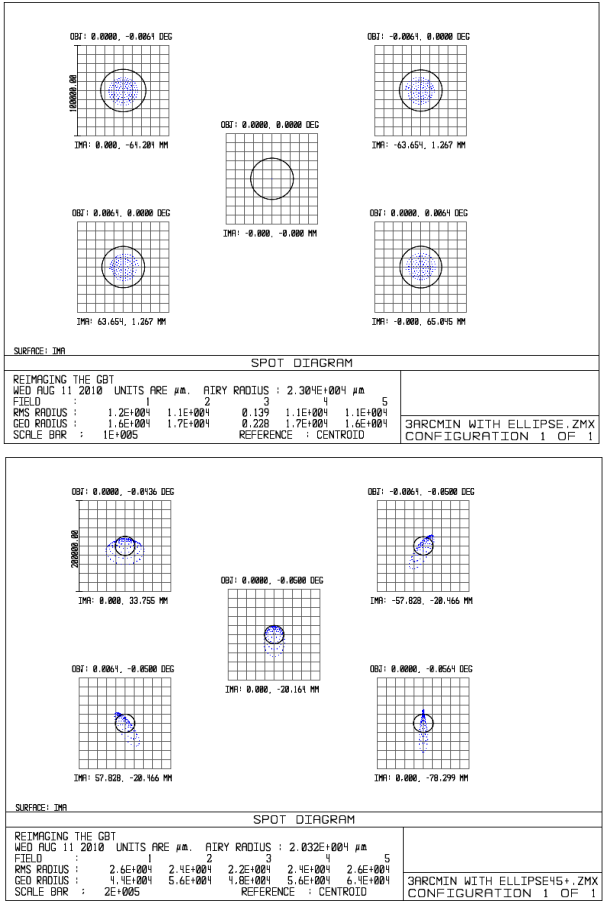

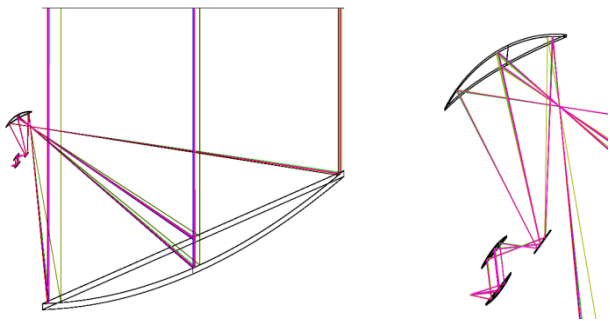

**Figure 5** *Upper*: Spot diagrams of a 23" field of view in the focal center using a flat chopper and an ellipse reflector mirror. *Bottom*: Spot diagrams of a 23" field of view at 3 arcmin away from the focal center.

**Figure 6** Optical arrangement using a flat chopper and two face-to-face poorly-aligned paraboloid reflectors.

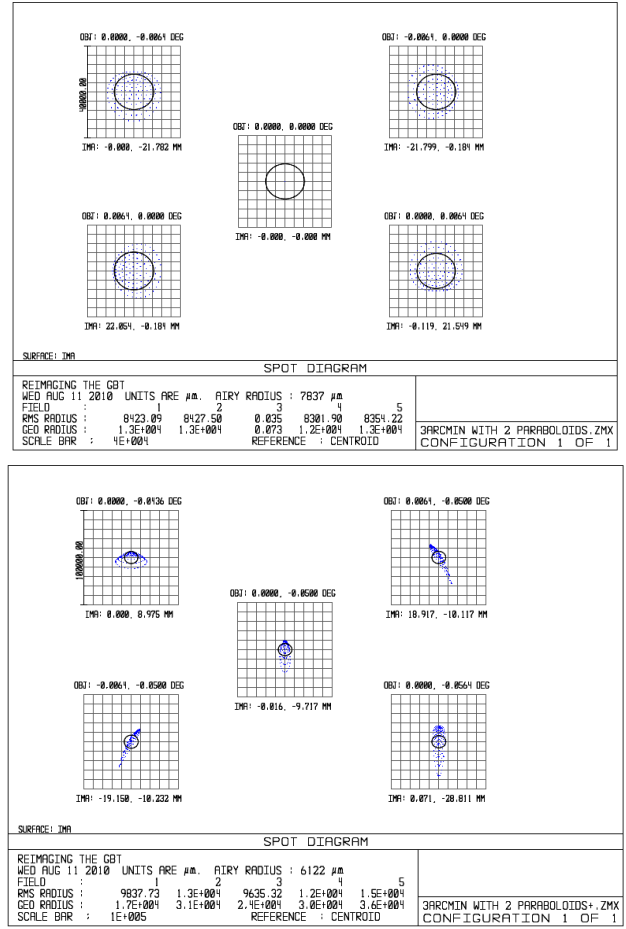

**Figure 7** *Upper*: Spot diagrams of a 23" field of view in the focal center using a flat chopper and a pair of poorly-aligned parabiodal reflector mirrors. *Bottom*: Spot diagrams of a 23" field of view at 3 arcmins away from the focal center

From the above discussion, the added image transferring system should be well-aligned with the focal plane of the GBT system. Figure 8 shows this improved flat chopper and a face-to–face, coma free, linear astigmatism free, paraboloid pair with their focal plane nearly parallel to that of the GBT optics. The imaging results are greatly improved. In Figure 8, the Y-Z plane is on the paper surface with Y towards the right hand direction and Z in the downward direction. The field scanning is done through the rotation of the flat chopper in the Y direction. When the flat mirror tilts about 2.6 degrees about the Y direction, the scanned field angle is 3 arcmin in the minus X direction. In the top diagram of Figure 9, the spot diagrams are at field angles (-0.0083 deg, 0); (0 ,-0.0083 deg); (0, 0); (0.0083 deg, 0); and (0, 0.0083 deg). The spot diagrams 3 arcmin away from the focal center are shown in the bottom of Figure 9. The spot diagrams are at the field angles of (-0.0583 deg, 0); (-0.05 deg, -0.0083 deg); (-0.05 deg, 0); (-0.05 deg, 0.0083 deg); and (-0.0417 deg, 0) respectively. When the beam is 3 arcmin away from the focal center, the rms spot radii with the MUSTANG device are 4.6 mm, 4.7 mm, 3.0 mm, 4.9 mm, and 4.7 mm respectively. These numbers are well below the Airy disk size of 7.72 mm. During chopping, the detector remains stationary at its fixed position. In this design the distance between two paraboloids can be adjusted without affecting the final image quality. This design can be used at the GBT focus to meet the image quality requirement in the MUSTANG chopping observation. The details of the scanner mirror parameters are all listed in Table 1.

|  | Flat chopper | Para 1 | Para 2 | Focus |
|---|---|---|---|---|
| Focal length |  | 2000 | 2000 |  |
| Tilt about X | -45 deg | 42 deg | 42 deg |  |
| Decenter about Y |  |  | -2650 mm |  |

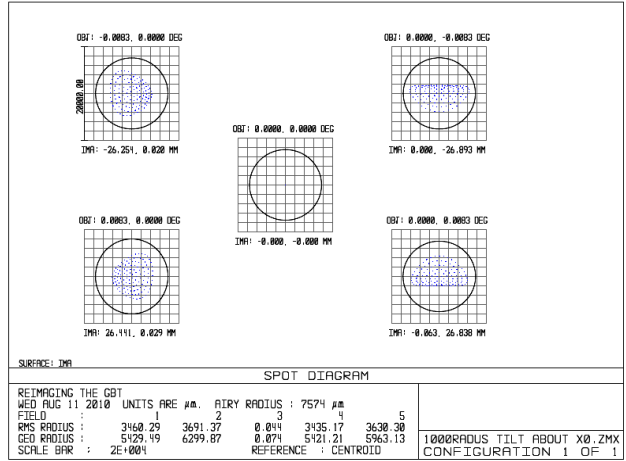

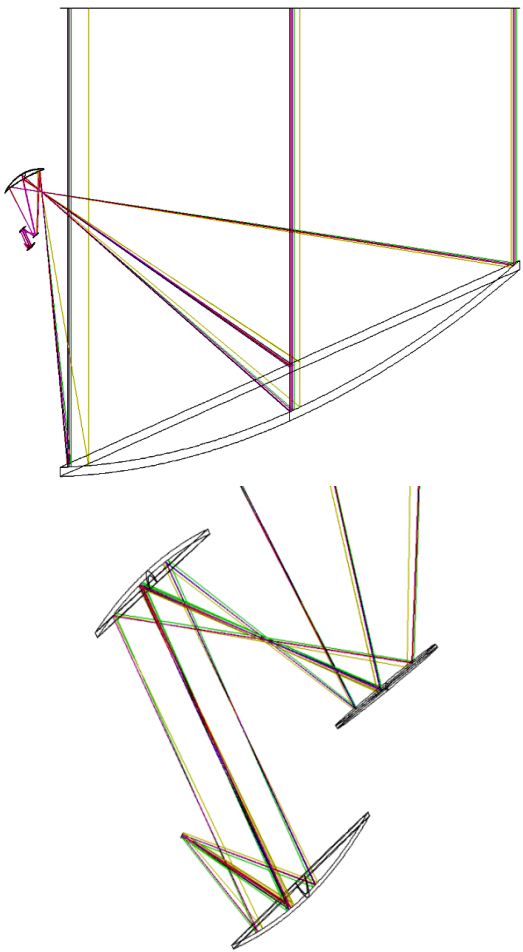

**Figure 8** Optical arrangement using a flat chopper and two face-to-face well-aligned paraboloid reflectors.

**Table 1** Position and parameter details of the scanner system.

|  | Flat chopper | Para 1 | Para 2 | Focus |
|---|---|---|---|---|
| Distance to previous surface | 1916 mm | 3416 | 3000 | 1500 |
| Diameter | 1000 mm | 1000 | 1000 |  |

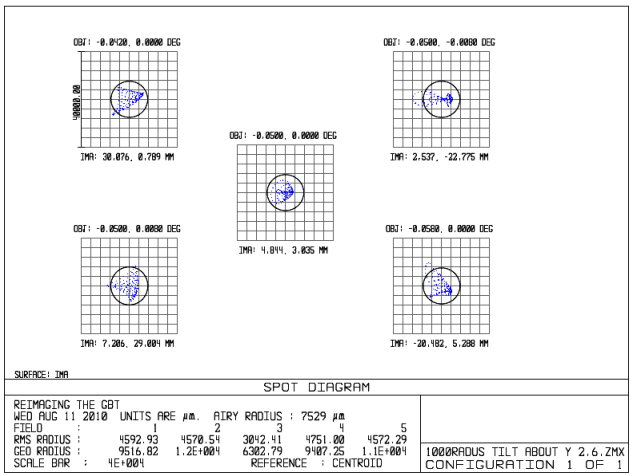

**Figure 9** *Upper*: Spot diagrams of a 23" field of view in the focal center using a flat chopper and a pair of well-aligned parabiodal reflector mirrors. *Bottom*: Spot diagrams of a 23" field of view at 3 arcmins away from the focal center.

## 4 Final remarks

Adding imaging transferring optics into an existing optimized off-axis system is difficult as the new systems produced may have an extremely small field of view. In this paper different schemes for scanning a small angular range of sky for the GBT are explored. Much attention was paid to retaining the image quality in the original GBT focal field. Finally, a field scanner with up to 3 arcmin field of view for the GBT is successfully designed which meets the image quality requirements of the MUSTANG device. In this design, the image transferring system is a pair well aligned face-to-face paraboloidals, which is an optimized coma free off-axis optical system. The added optical system adds very little aberrations in the final image. The field scanning is done by rotating a flat mirror with a tilting angle of about 1.3 degrees in the Y direction. The system can scan 3 arcmin sky areas. The rms spot size at 3 arcmin field angle is only a little more than 4 mm in diameter.

As this paper is still in preparation, a new bolometer camera (MUSTANG-2) is being designed. The new device has a larger field of view, about 1.5 arcmin in radius. The above designed system may not meet the requirement of MUSTANG-2. To design a suitable scanner for MUSTANG-2 requires further optimization. More attention has to be paid to reducing coma aberration which is inherited in the GBT off-axis optical system.

*We thank LIANG Ming for helpful discussions in this topic. The work of LI Yang and LI Xinnan were supported by the National Natural Sciences Foundation of China (No. 10978021). The National Radio Astronomy Observatory is a facility of the National Science Foundation operated under cooperative agreement by Associated Universities, Inc.*